# SILICENE NANORIBBONS ON AN INSULATING THIN FILM.

Khalid Quertite[1,2,3], Hanna Enriquez[1], Nicolas Trcera[2], Yongfeng Tong[2], Azzedine Bendounan[2], Andrew J. Mayne[1], Gérald Dujardin[1], Pierre Lagarde[2], Abdallah El kenz[3], Abdelilah Benyoussef[3,4], Yannick J. Dappe[5], Abdelkader Kara[6], and Hamid Oughaddou[1,7,*]

[1]Université Paris-Saclay, CNRS, Institut des Sciences Moléculaires d'Orsay (ISMO), Bât. 520, 91405 Orsay, France
[2]Synchrotron Soleil, L'Orme des Merisiers, Saint-Aubin, B.P. 48, 91192 Gif-sur-Yvette Cedex, France
[3]LaMCScI, Faculté des Sciences, Université Mohammed V - Agdal, 10100, Rabat, Morocco
[4]Hassan II Academy of Sciences and Technology, Rabat, Morocco
[5]Université Paris-Saclay, CEA, CNRS, SPEC, 91191 Gif-sur-Yvette Cedex, France
[6]Department of Physics, University of Central Florida, Orlando, FL 32816, USA
[7]Département de Physique, CY, Cergy Paris Université, 95031 Cergy-Pontoise Cedex, France

**Abstract**

**Silicene, a new two-dimensional (2D) material has attracted intense research because of the ubiquitous use of silicon in modern technology. However, producing free-standing silicene has proved to be a huge challenge. Until now, silicene could be synthesized only on metal surfaces where it naturally forms strong interactions with the metal substrate that modify its electronic properties. Here, we report the first experimental evidence of silicene sheet on an insulating NaCl thin film. This work represents a major breakthrough; for the study of the intrinsic properties of silicene, and by extension to other 2D materials that have so far only been grown on metal surfaces.**

\* Corresponding author: Hamid.oughaddou@universite-paris-saclay.fr

The discovery of graphene [1,2], with its unique physical properties has driven an intense research effort to find new 2D materials based on other elements [3,4]. Many have been discovered: boron nitride[5], silicene[6-12], germanene[13,14], phosphorene[15,16], stanene[17], and metal dichalcogenides[18-19]. The widespread use of silicon in modern technology makes silicene a particularly logical choice. A silicene-based field effect transistor operating at room temperature has been demonstrated recently [20]. Among the properties unique to 2D materials, silicene should show a quantum spin Hall effect [21], and giant magnetoresistance [22]. For historical reasons [9-10], the early investigations of silicene were performed almost exclusively on metal substrates by epitaxial growth of silicon atoms: Si/Ag[8-10], Si/Au[11-12], Si/ZrB$_2$[23], and Si/Ir[24]. However, a number of questions remained unanswered due to diverging evidence, crucially over the existence or not of Dirac fermions in silicene. Angular-resolved photoemission spectroscopy (ARPES) and scanning tunneling spectroscopy (STS) claimed to observe them in the case of silicene on silver[25,26], while other experiments indicated a strong interaction between silicene and the metallic surface[27,28], indicating that the observed linear dependence was due to an interface state of the silver. Using a metal substrate was obviously a hindrance to observing the intrinsic properties of silicene. This alone has motivated the exploration of other potential substrates to reduce the interaction with silicene. From this point of view, Alkali metal halides, such as NaCl, present an ideal alternative substrate, particularly in the form of a thin film, simply because the thin NaCl film behaves as a dielectric layer which decouples the electronic states of adsorbed species [29-31] from the metallic substrates. In current technology applications thick NaCl layers (50 nm) are used as a protective layer in flexible devices [32-34]. In the context of these technological developments, it is important to understand how an ultra-thin NaCl layer behaves if it is to be removed, especially since ultra-thin NaCl protects graphene, and is stable even in air, provided dry conditions are maintained (RH <30%) [35]. The literature shows that even in the ultra-thin limit, water will dissolve the NaCl layer. Early studies found rapid nanometer-scale step motion on NaCl above a critical relative humidity (RH) (40%) followed by deliquescence at 73%RH [36-38]. At the atomic scale, theory indicates that water extracts Na+ preferentially due to its higher affinity even though Cl- is favored energetically [39]. STM studies of the process of NaCl dissolution show the selective extraction individual Na+ ions with a water- functionalized tip [40]. Their recent theory study shows that the hydrated Na+ ions diffuse orders of magnitude faster than other hydrates [41]. Both temperature [40] and other electron sources [42] facilitate the process at the island edges.

Here, we present a comprehensive study of the growth of silicene on a thin NaCl film deposited on the Ag(110) surface. Using scanning tunneling microscopy (STM), low energy electron diffraction (LEED), x-ray photoemission spectroscopy (XPS), extended x-ray absorption fine structure (EXAFS) spectroscopy and density functional theory (DFT) calculations, we demonstrate that silicene forms an extended 2D structure on the thin NaCl film. The STM images show a highly ordered sheet with honeycomb-like structure. The XPS and EXAFS measurements reveal that the silicon atoms have a single chemical environment, and that the silicon environment is consistent with that of silicene. Finally, DFT calculations confirm the formation of weakly bound silicene composed of a densely packed array of nanoribbons on the NaCl film in line with the experimental results. This is the first clear evidence of the formation of silicene on an insulating thin film. The best growth conditions for obtaining large-area NaCl thin films on the Ag(110) surface has already been optimized successfully[43] (see Methods).

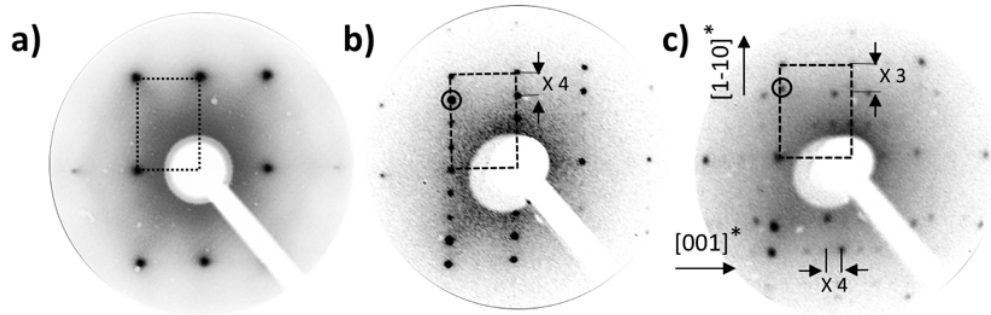

**Figure 1:** LEED patterns of (a) bare Ag(110) substrate, (b) after a deposition of ~ 1 ML of NaCl, (c) after the deposition of ~ 1 ML of silicon on NaCl/Ag(110). The Ag(110) unit cell, the x3 and x4 periodicities are indicated. The diffraction spot corresponding to NaCl film is highlighted by the small black circles in b and c. The LEED patterns were recorded at an energy of 70 eV.

Figure 1 presents the LEED patterns obtained on a) the bare Ag(110) surface, b) after deposition of ~ 1 ML of NaCl, and c) after the deposition of about 1 ML of silicon on the NaCl/Ag(110) surface, respectively. Figure 1-a shows a sharp (1x1) pattern characteristic of the clean Ag(110) surface. In Figure 1-b, the formation of the NaCl layer creates a (4x1) superstructure with respect to the Ag unit cell. In Figure 1c, the LEED pattern presents a (3x4) superstructure of the Si layer with respect to the Ag unit cell. In both Figures 1-b and 1-c, the underlying layers of Ag and NaCl/Ag, respectively, are visible through the top layer.

NaCl naturally forms islands on different metallic substrates [44-47]; the growth of NaCl on Ag(110) has already been optimized for this study[43]. With the substrate held at 410 K, large-area NaCl islands are obtained of more than 160 x 160 nm$^2$ in size[43]. Figure 2a shows a typical large-scale STM image of the thin NaCl film grown on Ag(110). The Ag(110) surface is covered by NaCl islands with very few bare areas of silver still remaining. The atomically resolved STM images shown in Figure 2b and 2c show that NaCl islands present a 1.15 nm x 0.40 nm rectangular unit cell which corresponds exactly to a (4x1) periodicity with respect to the Ag(110), in good agreement with the LEED observations. As expected, the atomic structure of the NaCl islands is no longer isotropic; the mono-domain Ag(110) crystal imposes an anisotropy on the NaCl growth. As a result, only a single domain orientation of the NaCl is observed as the LEED pattern clearly shows (Figure 1b). Furthermore, our detailed study of the NaCl growth on Ag(110)[43] showed that two configurations exist of the local density of states within the NaCl unit cell as a result of small variations in the local atomic structure as shown in Figure 2b and 2c. The orientation of the NaCl unit cell with respect to the substrate is shown in the atomically resolved STM image of the bare Ag(110) surface in Figure 2d confirming the agreement between the LEED and the STM observations.

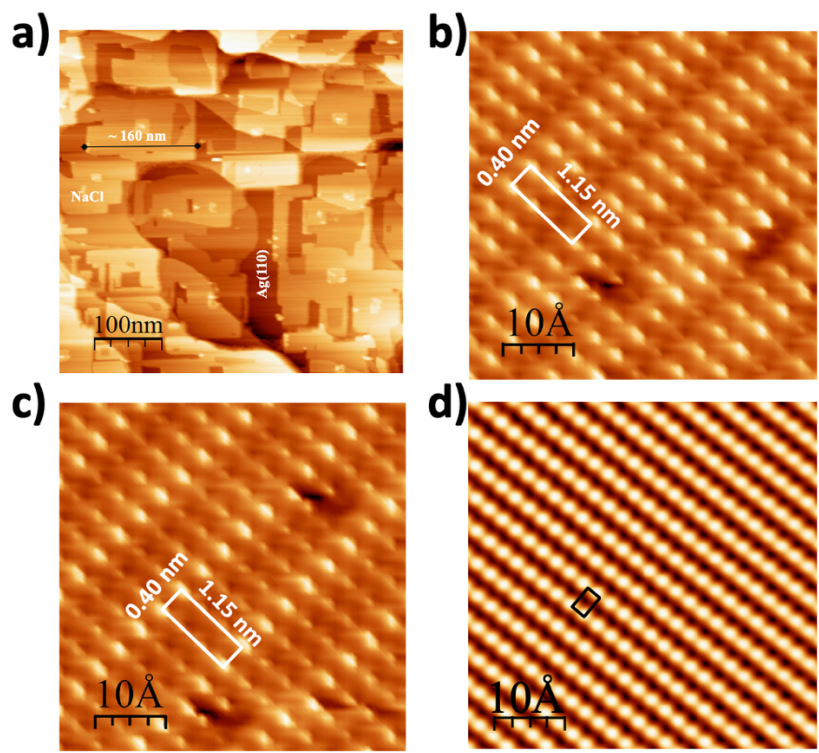

Figure 2: Occupied-state STM images corresponding to (a) 1ML NaCl film on Ag(110) (500x500 nm², U= -1.05 V, I= 0.7 nA), b) and c) atomically resolved STM images of NaCl islands (5x5nm², U= -1.0 V, I= 0.4 nA), the (4x1) unit cell is indicated by the white rectangle, d) atomically resolved STM image of the bare Ag(110) surface (5.5x5.5nm², U= -0.03V, I = 3.2 nA), the unit cell is indicated by the black rectangle.

Figure 3a shows an atomically resolved STM image *after* deposition of 0.2 ML of Silicon on top of the NaCl film. Small 2D structures of silicene (circled in red) are observed. The square structure of the NaCl (black square) and the (4x1) unit cell (black rectangle) are indicated. Following the deposition of the equivalent of 1 ML of silicon on the NaCl/Ag(110) surface, the STM topography in Figure 3b shows a Si layer that covers the entire surface of the NaCl islands. A zoom of Figure 3b (in Figure 3c) reveals a highly ordered honeycomb-like structure composed of parallel chains all oriented along the [1-10] direction. The line profiles recorded along the lines A and B in Figure 3c are shown in Figure 3d and 3e, respectively. In these line profiles, periodicities of 0.86 nm and 1.60 nm are found. These match the (3x4) LEED pattern because 0.86 nm is equivalent to 3 x $a_{Ag[1-10]}$ and 1.60 nm to 4 x $a_{Ag[001]}$, where $a_{Ag[1-10]}$ = 0.289 nm and $a_{Ag[001]}$ = 0.407 nm, respectively. In addition, the lateral distance between the two nearest protrusions is 0.45 nm (Figure 3d). This distance is larger than the expected first neighbor Si-Si distance (0.23 nm), indicating that only some of the atoms contribute to the Density of States in these STM images. Note that for silicon deposited directly on the Ag(110) surface, a nanoribbon structure forms with distinctly different (2x5) superstructure[48] (see Figure 3f).

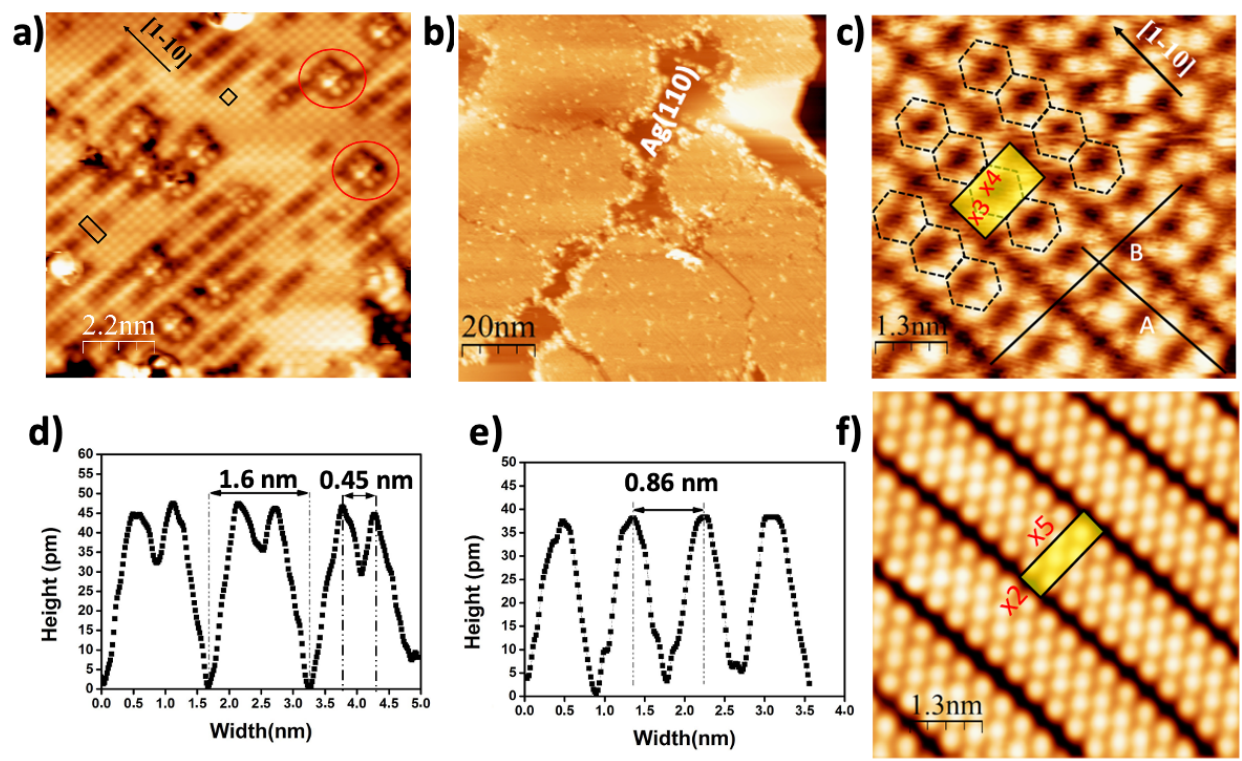

**Figure 3:** Occupied-state STM images corresponding to (a) atomic resolved STM image showing the first stages of growth of Si on NaCl islands deposited on Ag(110). The 2D-silicon structure is highlighted by red circles. For comparison, the (4x1) unit cell and the primitive (1x1) cell of the NaCl film are indicated by the black rectangle and square, respectively. (b) 1 ML of silicon deposited over the NaCl film (100 x 100 nm², U= -1V, I= 0,8 nA). (c) High-resolution STM image of the silicon layer (6.5 x 6.5 nm², U= -0.04 V, I= 2.6 nA). (d) and (e) show the line scans A and B in Figure 3c. From a separate experiment, (f) shows the silicene nanoribbons formed directly on the Ag(110) surface with the distinct 2x5 periodicity (6.6 x 6.6 nm², U= -0.03 V, I= 1.15 nA).

The electronic structure of the silicene layer can be understood further using XPS to probe the environment of the Si atoms within the silicene layer. The spectra corresponding to the Ag 3d, Si 2p, Cl 2p and Na 2s core levels were fitted with spin-orbit split Doniach-Sunjich (D-S) function [49] (see Methods). In Figure 4a, the Ag 3d core level spectra recorded with a photon energy of 700 eV and at normal emission (0° to the surface normal) are plotted for bare Ag(110) substrate. The Ag 3d core level was again recorded *after* deposition of the NaCl film, and then *after* deposition of Si on NaCl/Ag; these are shown in Figures 4b and 4c, respectively.

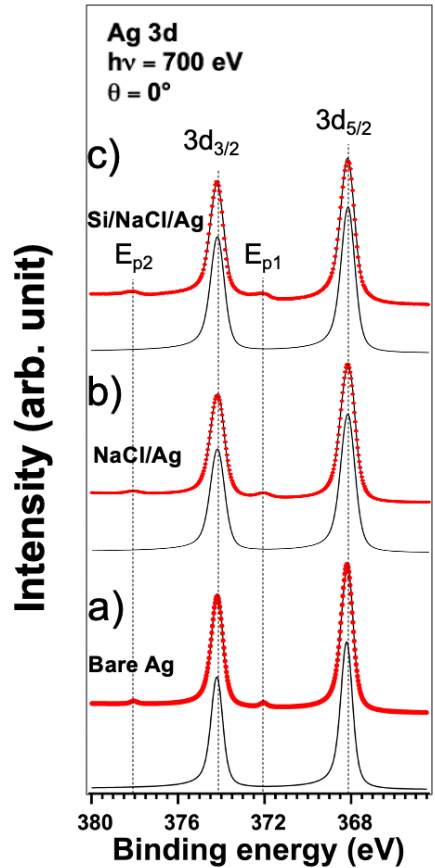

**Figure 4:** High-resolution core level spectra of Ag 3d recorded on bare Ag(110) surface (a) and after deposition of NaCl film (b) and after deposition on Silicon layer on NaCl (c). Dots correspond to data and black line overlapping the data corresponds to the best fit. For each spectrum, the component giving the best fit is indicated in black.

On the bare Ag(110) substrate, we observe two peaks in the XPS spectra at binding energies of 374.2 eV and 368.2 eV corresponding to the Ag $3d_{3/2}$ and $3d_{5/2}$ core level peaks. The peaks require a fit with only one spin–orbit split component, and are located energies similar to those reported in the literature [50]. The two small peaks located at $E_{P1}$ = 372.0 eV and $E_{P2}$ = 378.0 eV are the two dominant plasmon excitations of silver. After deposition of the NaCl film on Ag(110), the Ag 3d peaks are located at the same binding energies, and only one spin–orbit split component is required to fit the Ag 3d core level. This indicates that the interaction between Ag and NaCl is weak as expected. The two plasmon peaks of silver are still visible after deposition of NaCl indicating that the NaCl film has not affected the surface plasmons at the NaCl/Ag interface.

After deposition of the silicon layer on NaCl/Ag(110) only one spin–orbit split component is required to fit the Ag 3d core level. We observe that the deposition of silicon does not affect the plasmon peaks of silver. That they are still visible is an indication that Si atoms have not adsorbed on Ag surface, since these plasmons would be quenched following the adsorption of silicon on Ag(110) as has been shown previously[51]. It is clear that the silicene layer has grown on top of the NaCl layer.

A complete set of observations were obtained. In Figure 5a and 5b, the Cl 2p and Na 2s core level spectra were recorded, in normal emission (0°) at photon energies of 700 eV and 147 eV, respectively, *after* Si deposition on NaCl/Ag(110). The Cl and Na spectra require only one spin–orbit split component to fit the spectra. This indicates that Na and Cl ions have only one chemical environment (Na surrounded by Cl atoms and Cl atoms surrounded by Na). This again is further evidence for a weak interaction both between the NaCl film and Si as well between the NaCl film and the silver substrate. Finally, Figure 6c presents the Si 2p core level spectra of the silicon layer recorded in normal emission (0°) at a photon energy of 147 eV. The Si 2p spectrum requires only one spin-orbit split component to fit the spectrum indicating that the silicon atoms are all equivalent in having only one chemical environment. Given that Si atoms are adsorbed on top of the NaCl film and that the Si-NaCl interaction is weak. This contrasts clearly with previous XPS studies of silicon on silver surfaces that showed two distinct components resulting from the two Si-Si and Si-Ag environments [9,10,52].

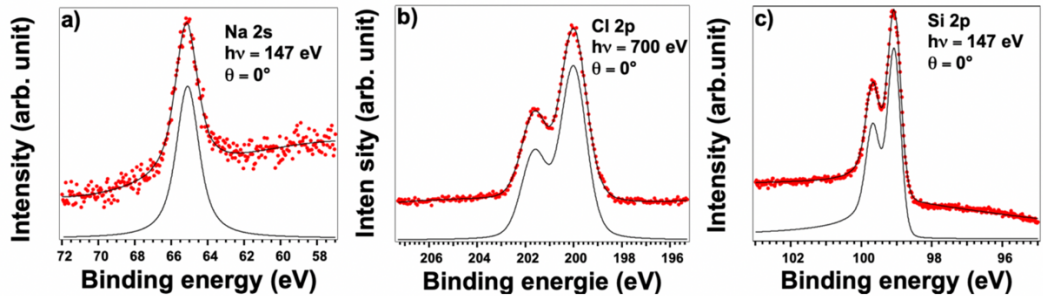

**Figure 5:** High-resolution core level spectra corresponding to: (a) Na 2s, (b) Cl 2p, recorded *after* deposition of 1 ML of NaCl on Ag(110), (c) Si 2p, recorded after deposition of 1 ML of silicon on top of the NaCl/Ag(110). All spectra are recorded at normal emission (0°). Dots correspond to the raw data and the overlapping black line corresponds to the best fit. For each spectrum, the component giving the best fit is indicated in black.

EXAFS measurements recorded at the Si K-edge provide complementary information on the local structure of the 2D silicon layer grown on NaCl/Ag(110) surface. EXAFS spectra give a precise determination of the interatomic distance between the excited atom (here Si) and its neighbors. Figure 6 shows the modulus and the imaginary part of the Fourier transform of $k^3X(k)$ for bulk silicon and for the silicon layer grown on NaCl/Ag(110) surface. We can observe that there is a high similarity between the two local structures. This indicates that the local structure around the Si atoms of the silicon layer is highly ordered. The peaks located at 1.92 Å and 3.4 Å in Figure 6a are assigned to the first and second interatomic distances in the bulk silicon crystal [53]. These values are shifted with respect to the expected lattice parameters of 2.35 Å and 3.83 Å due to the well-known effect of the phase shift between the absorbing atom and the backscattering one [54]. The first and second interatomic Si-Si distances from the EXAFS spectra are very close to the values reported for silicene grown directly on Ag(110) or Ag(111) surfaces[53].

Moreover, it is clear by comparing the imaginary part of the Si layer from this study (Figure 6b) with the Si bulk; only Si atoms are in the immediate vicinity of the excited atom (here the Si atom). It is interesting to note that in Ref 52, an interatomic distance of 2.7 Å was attributed to the Si-Ag bond [53]. In our study, this particular distance is not observed indicating that Si-Ag bonds are absent. Again, this shows that the Si atoms are located on top of NaCl film, not directly on the Ag surface. Taken together all the information from the STM, XPS and EXAFS studies, we find that silicon atoms adsorb on the ultra-thin NaCl film to form a 2D layer with a local structure similar to silicene that is decoupled electronically from the silver substrate.

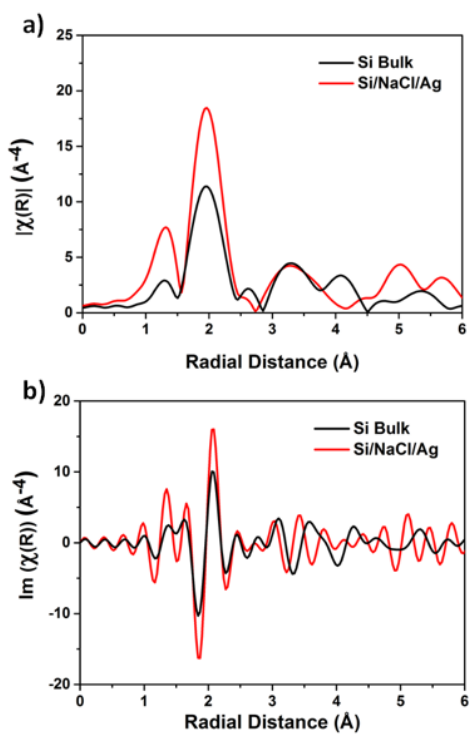

**Figure 6:** (a) modulus and (b) imaginary parts of the Fourier transforms of bulk silicon and 2D Si sheet deposited on NaCl/Ag(110).

To support the experimental findings, a computational study was performed using Density Functional Theory (DFT) (see Methods). From the LEED pattern shown in Figure 1b, along the close-packed direction, 3 Na-Na (or Cl-Cl) distances correspond to 4 Ag-Ag distances. Since the periodicity along this direction was x3 in the LEED after deposition of silicon (see Figure 1c), we have chosen a slab of 12 Ag-Ag lengths along this direction. As the NaCl and silicene layers form x1 and x4 periodicities along the direction perpendicular to the channels respectively (see Figure 1b and 1c), a periodicity x4 was adopted along this direction. The corresponding silicene structure which fulfills all these conditions is a silicene nanoribbon aligned along the long direction of the 12 x 4 slab. However, the edges of the nanoribbon have dangling bonds that strongly influence the structural properties of the nanoribbon. Two calculations were performed, one with unsaturated dangling bonds, and the second where the dangling bonds were saturated with hydrogen. We observed that hydrogen not only passivates the dangling bonds, but that the nanoribbon structure is stabilized with respect to that without hydrogen-passivated edges. We then optimized this hydrogen-terminated nanoribbon structure using the DFT calculations. This approach is fully justified both experimentally and theoretically. Several studies have shown that graphitic surfaces can have H-terminated step edges [55,56]. Theoretical calculations have also investigated H-terminated graphene and silicene ribbon structures [57,58]. Under UHV conditions, residual hydrogen is present which reacts with semiconductor surfaces (by cracking on the ion gauge filament) even at very low pressure ($10^{-11}$ mbar) [59].

Figure 7 shows the top and side views of the relaxed system. After full relaxation of the system, we found an interlayer distance of 0.33 nm between the silicene sheet and the NaCl layer.  The silicene sheet is much more corrugated than in the gas phase with a maximum corrugation of 0.67 Å. Note that even after having chosen a 12x4 cell of silver for the calculation, the Si presents a (3x4) structure with respect to the silver in good agreement with STM and LEED results. The (1x1) and the (3x4) cells are indicated in Figure 7. The first nearest neighbor Si-Si average distance is found to be 0.237 nm in good agreement with the EXAFS results and with the values reported previously for Silicene/Ag[53]. The 0.33 nm interlayer distance between the silicene and the NaCl is similar to that found in graphite and bi-layer graphene[60]. Taking into consideration the adsorption energy values of around 50 meV/atom as well, is strong evidence that silicene is held on the substrate by weak electrostatic van der Waals forces.

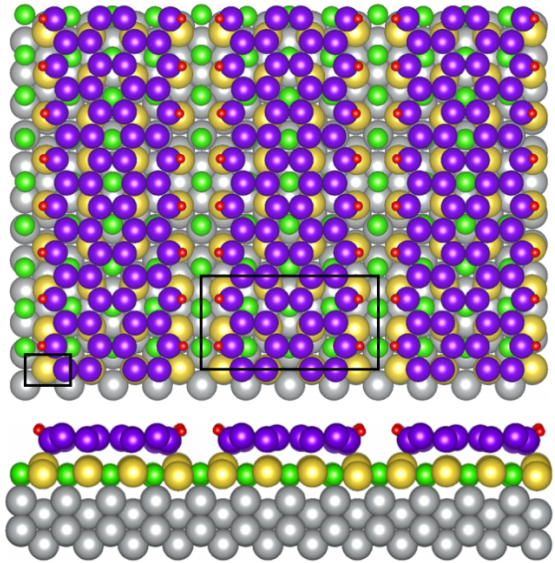

**Figure 7:** Top view (up) and side view (down) of silicene nanoribbons on NaCl/Ag(110), after full relaxation. Ag, Cl, Na, H, and Si atoms are in grey, green, yellow, red and blue purple. Top view (up) and side view (down). The (1x1) and the (3x4) cells are indicated by the small and large rectangles, respectively.

**Methods**

The experiments were performed in ultra-high vacuum apparatus with pressure in the $10^{-11}$ mbar range, equipped with special tools for surface preparation and characterization: an ion gun for surface sputtering, Low Energy Electron Diffraction (LEED), Auger electron spectroscopy (AES) and a commercial Omicron LT-STM. The growth of NaCl and Silicon was controlled systematically by Auger and LEED measurements.

Surface preparation

The Ag(110) commercial crystal with 99.99999% purity was cleaned by multiple cycles of sputtering (650 eV Ar+ ions, P ~ $10^{-5}$ mbar) followed by annealing at 500 °C. NaCl deposition was done using a Knudsen cell heated at 520 °C. The Ag substrate was kept at 140°C during NaCl deposition to grow large NaCl islands [43]. Silicon was evaporated (~ 0.02 Si ML/min) by direct current heating of a Si wafer to about 1200 °C onto NaCl/Ag(110) held at 140 °C. A post-annealing was performed at 200 °C in order to improve the diffusion of silicon. All the STM images where recorded at 78 K and Nanotec WSxM software [61] was used to process and analyze the data.

Spectroscopy

The XPS and EXAFS experiments were carried out at TEMPO and LUCIA [62] beam-lines respectively at SOLEIL Synchrotron facility in France. The same Ag(110) crystal was used for all the experiments. The core level spectra were adjusted with spin-orbit split Doniach-Sunjich (D-S) function. The Ag 3d was fitted with a 140 meV eV Gaussian profile and a 280 meV Lorentzian profile, the spin-orbit splitting is 6 eV, and the branching ratio is 0.78. The best fit for Cl 2p was obtained with a 0.83 eV Gaussian profile and a 0.12 eV Lorentzian profile, the spin-orbit splitting is 1.63 eV, and the branching ratio is 0.49. For Na 2s, the best fit is obtained with 0.53 eV Gaussian profile and a 0.55 eV Lorentzian profile, while for the Si 2p spectra, the best fit was obtained with a 0.255 eV Gaussian profile, a 0.07 eV Lorentzian profile, the spin-orbit splitting is 0.61 eV, and the branching ratio is 0.48. The background of each spectrum was fitted with polynomic of second order. The EXAFS spectra were collected in total electron yield mode. The full analysis of the experimental data has been conducted with Athena software from the DEMETER package [63]. A spline curve has been used to remove the background oscillations. The Fourier transforms (FT) have been applied to $K^3 \cdot \chi(k)$ raw data, with a Kaiser-Bessel apodization window spanning from 2 to 9.5 Å$^{-1}$.

Structure computation

The computational study has been performed using Density Functional Theory (DFT) as implemented in the Fireball code [64]. This code uses the local density approximation (LDA) for the exchange and correlation functional according to the McWeda approach [65] and a combination of atomic-like orbitals as basis set [66]. Basis sets of sp$^3$d$^5$ for Ag and Cl, sp$^3$ for C and Si and s for H and Na were used with cutoff radii (in atomic units) s = 4.5, p = 5.5, d = 4.0 (Ag), s = 3.9, p = 4.4, d = 5.4 (Cl), s = 4.5, p = 4.5 (C), s = 4.8, p = 5.4 (Si), s = 4.1 (H) and s = 8.0 (Na). These basis sets have been already used successfully in the study of graphene and two-dimensional materials [67,68]. We have considered a Ag(110) slab of 5 layers containing 12 atoms each in a 3x4 structure, with a NaCl monolayer on top plus a hydrogenated silicene nanoribbon. Calculations have been performed using a set of 32 k-points in the first Brillouin zone, and structures have been optimized until the forces were lower than 0.1 eV/Å.

**Notes**

The authors declare no competing financial interests.

**Acknowledgements**

This work is supported by a public grant overseen by the French National Research Agency (ANR) as part of the "Investissements d'Avenir" program (Labex NanoSaclay, reference: ANR-10-LABX-0035).